\documentclass[preprint,proceedings]{rmaa}

\usepackage{paralist}

\usepackage{psfrag,color}

\SetYear{2005}
\SetConfTitle{XI Latin American Regional IAU Meeting}

\title{Evidence for rotation-induced mixing in evolved intermediate mass stars} 

\author{
  R. Smiljanic,\altaffilmark{1} 
  B. Barbuy,\altaffilmark{1}
  J.R. De Medeiros,\altaffilmark{2}
  and A. Maeder\altaffilmark{3}}

\altaffiltext{1}{Universidade de S\~ao Paulo, Brazil.}
\altaffiltext{2}{Universidade Federal do Rio Grande do Norte, Brazil.}
\altaffiltext{3}{Geneva Observatory, Switzerland.}

\shortauthor{Smiljanic et al.}
\shorttitle{Rotatio-induced mixing in intermediate mass stars}

\fulladdresses{
\item Beatriz Barbuy and Rodolfo Smiljanic: Universidade de S\~ao Paulo, IAG, Rua do Mat\~ao
1226, Cidade Universit\'aria, 05508-900, S\~ao Paulo - SP, Brazil
  (\email{[barbuy, rodolfo]@astro.iag.usp.br}).
\item Andr\'e Maeder: Geneva Observatory, 1290 Sauverny, Switzerland (\email{Andre.Maeder@obs.unige.ch}).
\item Jos\'e Renan de Medeiros: Universidade Federal do Rio Grande do Norte, Dep. de F\'{\i}sica, Campus Universit\'ario, 59072-970, Natal, RN, Brazil (\email{renan@dfte.ufrn.br}).
}

\listofauthors{R. Smiljanic, B. Barbuy, J.R. De Medeiros \& A. Maeder}
\indexauthor{Smiljanic, R.}
\indexauthor{Barbuy, B.}
\indexauthor{De Medeiros, J.R.}
\indexauthor{Maeder, A.}

\abstract{Many observational results seem to indicate more efficient mixing processes in intermediate mass stars (5-20 M$_{\odot}$) than the expected by the standard models. These processes are usually thought to be caused by stellar rotation. Our recent analysis of 19 evolved intermediate mass stars has found them to display different efficiencies of internal mixing. The comparison of these results, and others from the literature, with rotating and non-rotating stellar evolutionary models led us to find, for the first time, an important correlation between stellar mass and the [N/C] ratio; the kind of correlation expected to be produced by a rotation-induced mixing.}

\addkeyword{Stars: Evolution}
\addkeyword{Stars: Rotation}
\addkeyword{Stars: Supergiants}
\begin{document}
\maketitle

\section{Introduction}
\label{sec:intro}
\par Intermediate mass stars burn hydrogen during the main sequence (MS) via the CNO cycle, converting almost all central C$^{12}$ into N$^{14}$. When the star evolves to the red giant branch (RGB) it suffers the first dredge-up, a deep convective layer that brings nuclear processed material to the surface. The photospheric abundances of C and N are then altered (C is reduced and N is increased). Although the theoretical models predict the first dredge-up to be the first important mixing event, observational evidence seems to point towards a more complex scenario. 
\par There is evidence of He and N overabundances in O and B stars and also for boron depletion in main sequence B stars. A-type supergiants of the Galaxy and of the SMC also show evidence for mixing before the first dredge-up. F type supergiants have been found with abundances higher than expected, which indicates a more efficient mixing, but also with non-modified and slightly modified abundances (see references in Smiljanic et al.\@ 2006). These observations suggest both a mixing process during the MS and a more efficient mixing than the expected solely due to the first dredge-up on the RGB. Neither are expected from the standard non-rotating models. 
\par The inclusion of rotation in the models seems to be able to reproduce these behaviors, at least qualitatively. Effects induced by rotation, such as meridional circulation (Maeder \& Zahn 1998) and shear turbulence (Maeder 1997; Mathis \& Zahn 2004) act in transporting and mixing the chemical elements. Thus, the required additional mixing appears naturally when rotation is taken into account. Even a mixing event during the MS should happen if the star rotates fast enough.

\section{Discussion}
\label{sec:disc}

We have determined atmospheric parameters, masses, and CNO abundances for a sample of 19 evolved intermediate mass stars (see details in Smiljanic et al.\@ 2006). The spectra were obtained with the FEROS spectrograph at the ESO 1.52m telescope at La Silla (Chile). They have a resolving power of R=48,000 and S/N higher than 200.
\begin{figure}\centering
  \includegraphics[width=0.8\columnwidth]{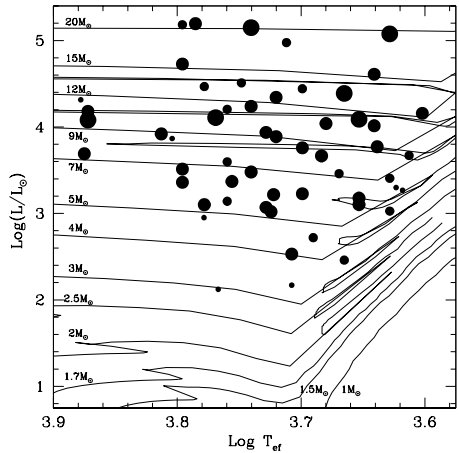}
  \caption{HR diagram showing our sample together with the stars from Luck \& Lambert (1985) and Barbuy et al.\@ (1996), overplotted on tracks from Schaller et al.\@ (1992). The circles are proportional to the [N/C] ratio in the following order (from smaller to larger): [N/C] $<$ +0.52, +0.52 $<$ [N/C] $<$ +0.92, +0.92 $<$ [N/C] $<$ +1.44 and [N/C] $>$ 1.44.}
  \label{fig:hr}
\end{figure}
\par We found signs of internal mixing in fifteen stars, with a mean [N/C] ratio of +0.95 dex. Only five of them have abundances in agreement with the non-rotating models by Meynet \& Maeder (2000) ([N/C] = +0.72 dex). One star is less mixed than that but in marginal agreement with the non-rotating models by Schaller et al.\@ (1992) ([N/C] = +0.60 dex). 
\par The mixing in the other sample stars was more efficient. The rotating models by Meynet \& Maeder (2000) show a better, but not pefect, agreement with the observed abundances. Comparing our results and the ones from Luck \& Lambert (1985) and Barbuy et al.\@ (1996) with the predictions, we found some stars more massive than 8M$_{\odot}$ to be more mixed than expected for the rotating models. We also found 5M$_{\odot}$ stars as mixed as that expected for 12M$_{\odot}$ stars.
\par Figure \ref{fig:hr} shows the distribution of the stars along the HR diagram, divided according to their abundances. The different groups of stars overlap because of the blue loops. In figure \ref{fig:hr} we clearly identify a group of fully mixed 5M$_{\odot}$ stars in a region where blue loops are not expected. It is not clear whether this is due to incorrect masses, to an underestimated extent of the loops or observational errors.
\par
\begin{figure}[!t]\centering
  \includegraphics[width=0.8\columnwidth]{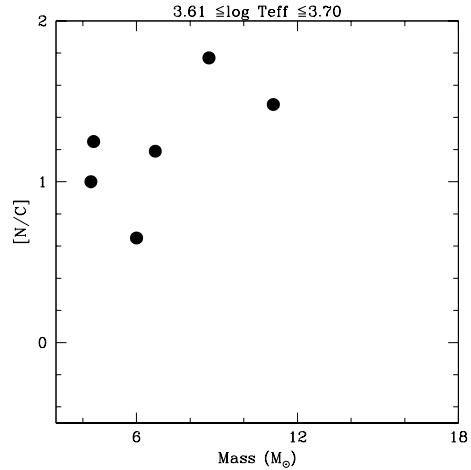}
  \caption{Plot of [N/C] vs. stellar mass in the interval between log T$_{\rm{eff}}$ = 3.61 and log T$_{\rm{eff}}$ = 3.70 where a clear correlation can be seen. Only the stars with $vsini$ $>$ 6 km s$^{-1}$ are plotted.}
  \label{fig:ncmass}
\end{figure}

 The [N/C] values depend on many factors such as stellar mass, initial $vsini$ and evolutionary state. In order to disentangle these effects, [N/C] vs. mass was plotted for the stars divided in intervals of temperature. A correlation is seen in the interval between log T$_{\rm{eff}}$ = 3.61 and 3.70, the red supergiants, for stars with $vsini$ $>$ 6 km s$^{-1}$ (Fig \ref{fig:ncmass}), indicating \adjustfinalcols a large increase in [N/C] with increasing mass.
\par This is the first time that such a relation is obtained and it may represent a new and important constraint for the evolutionary models that account for rotation. However, we have still to keep in mind that it is defined by only a few points and that there seems to be a large scatter. The scatter, however, is probably due to the fact that a large number of stars will reach the RGB with $vsini$ $>$ 6 km s$^{-1}$. Further work in extending the sample and confirming the correlation and the scatter is still needed.

\acknowledgements

The authors acknowledge support from the brazilian agencies CAPES, CNPq, FAPESP and FAPERN.

\end{document}